\documentstyle[11pt,psfig]{article}

\newcommand{\be}{\begin{equation}}
\newcommand{\bel}{\begin{equation}\label}
\newcommand{\ee}{\end{equation}}
\newcommand{\bea}{\begin{eqnarray}}
\newcommand{\eea}{\end{eqnarray}}

\newcommand{\eps}{\epsilon}
\newcommand{\ga}{\gamma}
\newcommand{\om}{\omega}

\renewcommand{\/}{\over}

\renewcommand{\[}{\left [}

\author{
Holger Schanz and Bernd Esser \\[5mm]
Institut f\"ur Physik,
Humboldt~- Universit\"at\\
Invalidenstr.110,
10 099 Berlin, Germany\\[1mm]
{ \footnotesize Tel./Fax: +49-30-2803-236/238},
{ \footnotesize email: schanz@summa.physik.hu-berlin.de}
}
\title{ 
Nonadiabatic couplings and incipience of quantum chaos}

\begin{document}

\maketitle
\date{January 19, 1996}

\begin{abstract}
\noindent
The quantization of the electronic two site system interacting with a
vibration is considered by using as the integrable reference system
the decoupled oscillators resulting from the adiabatic approximation.
A specific Bloch projection method is applied which demonstrates how
besides some regular regions in the fine structure of the spectrum and
the associated eigenvectors irregularities appear when passing from
the low to the high coupling case. At the same time even for strong
coupling some of the regular structure of the spectrum rooted in the
adiabatic potentials is kept intact justifying the classification of
this situation as incipience of quantum chaos.
\end{abstract}

\pagebreak
\section{Introduction}
Nonadiabatic couplings are essential to derive from the 
basic idea of the Born~- Oppenheimer (BO) approximation a 
rigorous procedure and were introduced already by Born itself in 
order to put this approximation into a complete scheme \cite{BH}. 
As has recently become clear these couplings can be the source of 
nonintegrability and produce chaotic behaviour in the dynamics of 
systems treated in the mixed quantum~- classical description 
(see e.~g.~\cite{BK,BE}). In this paper the manifestations of 
these couplings in the structure of the energy spectrum and the eigenvectors are 
considered for a fully quantized electronic~- vibronic coupled 
system which displays chaotic dynamics in the mixed description 
\cite{ES}. The electronic subsystem is a two site system (dimer) 
represented by a two state basis, whereas for the vibronic 
subsystem a large number of basis states has been taken into 
account in order to meet the basic condition behind the BO 
approximation for dividing the total system into fast 
(electronic) and slow (vibronic) subsystems. Furthermore 
this is the situation when one expects the best correspondence
when passing for the slow subsystem from the classical approximation to the quantum 
description. 
This last step completes the quantization according 
to the BO scheme and is of general interest 
in the context of the problem
how chaos in systems with a mixed quantum~- classical 
description is reflected in their spectrum after full quantization 
is performed \cite{BEII}.

The problem, which signatures of chaos the eigenvalue spectrum displays
in systems with a finite number of electronic quantum states
(representable as spin states) coupled to a vibration (boson field) has
already been addressed in the early papers on quantum chaos (see
\cite{GH,K}).  The result of this research was that despite some
irregularities in the spectrum the level spacing statistics is mostly
regular. Here we demonstrate for one explicit system of this class,
namely the excitonic~- vibronic coupled dimer (for details of the model
see \cite{ES}), that in fact in this kind of systems one can follow how
the irregularities in the eigenvalue spectrum and the associated eigenstates 
appear through patterns of growing complexity when a
characteristic coupling parameter is increased. Hence these systems have
to be placed between systems with regular and completely irregular
spectra. The appearance of irregularities in the eigenvalue spectrum or
incipience of quantum chaos was recently reported for the spectrum of
the spin~- boson Hamiltonian with the Jaynes~- Cummings model as the
integrable reference system \cite{CCM}. Here we use a different approach
focussing on the properties of the spectrum for oscillator frequencies
small compared to the electronic transfer matrix element, because we
consider as the integrable reference system the decoupled oscillators
resulting from the adiabatic approximation. Correspondingly we select a
parameter region very much different from \cite{CCM}. This
approach deserves particular attention since the widespread use of the
adiabatic approximation makes the problem, how the spectrum is modified
when a rigorous scheme including nonadiabatic couplings 
is applied and which are the differences in this
respect between regular and chaotic systems very significant. Using the
the excitonic~- vibronic coupled dimer as an example we show that for
the low coupling case the overall structure of the spectrum of the
electronic~- vibronic system can easily be understood from the adiabatic
approximation. For higher coupling it is possible to observe how the
mixing of the adiabatic sequences of levels results in the appearance of
irregularities in the spectrum.  In particular, from our approach it is
evident that this "birth" of irregularities, or incipience of quantum
chaos, is induced by the nonadiabatic couplings in the region of the
energetic overlap of the adiabatic potentials. 

\section{The Model}
The electronic two site system coupled to a vibration is 
represented in the form
\bel{hamiltonian}
H =  \pmatrix{\eps + {\ga\/\sqrt{2}}\, q + U(q) & V \cr 
V & \eps - {\ga\/\sqrt{2}}\, q + U(q)} + {\hat T}\, {\mathbf I}           
\ee
where $\eps$ is the electronic site energy (symmetric dimer), $V$ the transfer 
matrix element between the sites, $q$ the oscillator coordinate, 
$U(q)$ its potential energy and $\ga$ the coupling constant. 
$\hat T$ and $\mathbf I$ are the kinetic energy of the oscillator
and the identity operator in a two state basis respectively. 
We set $\hbar = 1$ and $\eps = 0$.

The separate diagonalization of the first term in (\ref{hamiltonian}) 
yields as eigenvalues the adiabatic potentials
\bel{adpot}
U^\pm_{\mathrm ad}(q) =   \pm \sqrt{V^2 + {{\ga^2}\/2} q^2} + U(q),  
\ee
which display the splitting into the bonding (-) and antibonding 
(+) branches. In the adiabatic approximation these branches are 
considered as independent and the action of $\hat T$ is restricted to 
the corresponding subspaces.
The full action of $\hat T$ includes the nonadiabatic couplings
resulting in a mixing of the adiabatic eigenstates.

We investigated the fine structure of the spectrum and the the 
associated eigenstates of the 
Hamiltonian (\ref{hamiltonian}) 
including the nonadiabatic couplings for a harmonic potential 
$U(q)= (\om^2 /2) q^2$. The problem depends on two dimensionless 
parameters which were chosen as $r={\om\/2V}$ and $p={\ga^2\/2V\om^2}$.
The parameter $p$ 
represents the coupling in these units. An important 
point is that $p$ appears in form of the product $\sqrt{p}r$ in the 
off diagonal elements of the Hamiltonian matrix in the adiabatic basis, 
i.~e.~the nonadiabatic mixing is governed by the effective 
coupling $\sqrt{p}r$. Below $p$ is used as the second independent parameter, 
however, because this parameter controls the bifurcation of the 
stationary states associated with the bonding branch $U^-_{\mathrm ad}(q)$ and 
the dynamics of the model in the mixed quantum~- classical 
description \cite{ES}. 

  In order to establish the influence of the nonadiabatic couplings 
  the eigenvalue spectrum in
  the adiabatic approximation has to be found.
For the determination of the adiabatic branches we use that in 
the region of highly excited oscillator states, i.~e.~large 
quantum numbers, the approximation 
\bel{apadpot}
 U_{\mathrm {ad}}^{\pm}(q) \sim \pm {\ga\/\sqrt{2}}|q| + U(q) =  
-{\om^2\/2}q_0^2 +{\om^2\/2}(|q|\pm q_0)^2,\hspace*{5mm} 
q_0 ={\ga\/\sqrt{2}\om^2}
\ee
can be used in (\ref{adpot}),
which captures both the splitting of the ground state into two 
degenerate minima and the asymptotic dependences on $q$ for large $q$. 
The branch $U_{\mathrm {ad}}^-$ represents the textbook problem of the
double oscillator \cite{Merz}. Its solution by parabolic cylinder
functions can easily be modified to include the branch $U_{\mathrm {ad}}^+$
as well. Using the asymptotic expansion of the parabolic cylinder functions
one obtains 
the following expression for the spacings between adjacent levels with the same parity, 
$\Delta E_i = E_{i+1} - E_i$ in the 
adiabatic approximation and for large $E_i$
\bel{apspacing}
\Delta E_i^\pm \sim {2 r \/ 1 \mp  {1 / \pi \sqrt{p/E_i^\pm}}    },
\ee
where $\pm$ ($\mp$ in the denominator) refer to the upper / 
lower part of the adiabatic potentials (\ref{adpot}), (3). 

\section{Analysis of the Numerical Results}
We performed a numerical diagonalization of the Hamiltonian (1) in a finite
electronic~- oscillator basis including the nonadiabatic couplings.
The adiabatic case of small oscillator frequencies compared to 
the electronic matrix element $V$ was considered, i.~e.~$r\ll 1$. Then, 
in order to reach the region of energetic 
overlap of the two adiabatic level sequences where the 
nonadiabatic mixing occurs, a relatively 
large number of vibrational basis states has to be taken into 
account (in our case of the order of $10^3$). 

The Hamiltonian (\ref{hamiltonian}) commutes with the total 
parity operator 
which exchanges the sites of the dimer and reflects the oscillator
coordinate. Correspondingly the eigenstates are characterized by their parity.
Below the results for the positive parity states are presented, 
the results for negative parity are similar.

Analyzing the eigenvalues and associated eigenvectors we 
applied a specific projection method projecting the 
eigenstates onto the Bloch sphere of the electronic subsystem. 
This projection is performed by calculating for each eigenstate 
the expression $x_i^{(\nu)} = c^{(i)*}_{\nu,1} c_{\nu,2}^{(i)} + c_{\nu,1}^{(i)} c_{\nu,2}^{(i)*}$, where the $c_{\nu,s}^{(i)}$ are the 
expansion coefficients of the eigenvector $i$ in the combined 
electronic~- vibronic basis, the indices $s=1,2$ refer to the 
electronic two site basis and $\nu=1,\dots N$ to the vibrational basis 
states. $N$ is the total number of vibrational basis functions, 
which are taken as harmonic oscillator eigenstates. Then the sum 
$x_i = \sum_\nu x^{(\nu)}_i$  over all vibrational states is calculated which 
gives the average Bloch variable $x_i$ associated with each 
eigenstate. $x$ is a useful quantity to distinguish 
between the mainly bonding or antibonding nature of the states in 
the electronic subspace associated with the two branches of the 
adiabatic potential (\ref{adpot}) (for the details of the Bloch 
representation of the dimer see \cite{ES}). We recall that $x=+1$ and 
$x=-1$ for a pure bonding or antibonding state, respectively. Below 
we use the sign of $x$ as an indicator of the mainly bonding or 
antibonding nature of the eigenstates.

A standard representation to resolve the fine structure of the
spectrum is the dependence of the energetic difference of two
subsequent eigenvalues $\Delta E_i = E_{i+1}-E_i$, i.~e.~the level
spacing, on the position of the eigenvalues $E_i$ itself. In Fig.~1(a)
such a dependence is shown for the dimensionless coupling strength
$p=5$ and $r=0.025$ (effective coupling $\sqrt{p}r\sim 0.06$).  In the energy
region below $E = 0.5$ only the lower adiabatic potential which has a
double well structure for the given parameters contributes to the
spectrum. Consequently, the spectrum is regular and the spacings
between consecutive energy levels show a smooth dependence on the
energy with a cusp marking the saddle point of the double well. For
energies above the minimum of the second adiabatic potential, in
addition to the points on the smooth curve in the $\Delta E_i(E_i)$
plot, which is still visible, we find points irregularly scattered
below this curve. However, this irregularity is merely due to a
superposition of the adiabatic level sequences which for this low
coupling case interact only weakly. For assigning the individual
levels to the two adiabatic ladders we have employed the projection of
the associated eigenvectors onto the Bloch sphere mentioned above. In
Fig.~1(b) the expectation value of the Bloch $x$~- variable and its
variance are displayed. It is evident that the eigenvectors can be
approximately divided into two groups which are the analogue to the
strands introduced in \cite{CCM}.  In Fig.~1(c) only the spacings
between eigenvalues belonging to the same group have been plotted. It
is obvious that the spectrum has been resolved into two smooth
dependences which are very close to the approximate level spacings of the 
adiabatic potentials (4) shown as full lines in Fig.~1(c).
We conclude, that the spectrum for this low coupling
case can be well understood in the adiabatic approximation.

In the set of figs.~2(a)-(c) the same type of dependences as in
figs.~1(a)-(c) are displayed for the parameters $p=2$ and $r=0.1$
giving $\sqrt{p}r\sim 0.14$ which is above the effective coupling of Fig.~1.
However, due to the higher value of $r$, in Fig.~2 a much larger energy
interval is required to display the same number of eigenstates.  
The functions (4) corresponding to the
adiabatic potentials (3) are again shown as lines in Fig.~2(c).  
It is seen
that for the given coupling value most of the spacings are still
reproduced by the integrable reference oscillators of the adiabatic
approximation. At the same time a substantial part of the data is
displaced from the adiabatic dependences indicating that there are
pairs of states with mainly bonding or antibonding nature completely
outside their adiabatic strands, i.~e.~due to the nonadiabatic mixing
the spectrum cannot be described as a superposition of two independent
sequences.

The effect of stronger mixing of the bonding and antibonding 
components in the eigenstates for a still higher 
coupling is evident from figs.~3(a)-(b), where the parameters $p=15$ and 
$r=0.1$, the effective coupling being $\sqrt{p}r\sim 0.39$, were used. The 
spacings of Fig.~3(a) show regular regions interrupted by 
regions with irregularly scattered data. The average Bloch 
variables have now merged into a band located around the $x=0$ 
line (Fig.~3(b)), i.~e. a reasonable distinction between the eigenstates as 
mainly bonding and antibonding is not possible any more (compare 
to Fig.~2(b) where this distinction can still be made) and there is no 
separation in the fine structure of the spectrum, such as shown 
in Fig.~2(c) possible. 
As in the spacings data there are regular 
regions interrupted by irregular parts in the transition 
between the upper and lower boundaries of the band of Bloch 
variables. The band of average Bloch variables is now entirely embedded into 
the variances which provide additional information about the 
stronger degree of mixing between the bonding and antibonding 
components of eigenstates, compare figs.~1(b)-3(b). This comparison 
demonstrates that with increasing coupling the characterization 
of the eigenstates by quantities other than the energy, such as 
the Bloch variables becomes increasingly meaningless.

Finally we turn to the examination of a representative high coupling
case. In Fig.~4 the dependences analogous to Fig.~3 are represented
for the parameter values $p=50$ and $r=0.1$, the effective coupling
being $\sqrt{p}r\sim 0.71$. Comparing Fig.~4(a) with the figs.~2(a) and 3(a) it is
seen that the lower boundary of the spacing data is shifted to higher
values, i.~e.~there is a clear tendency of repulsion between all the
pairs of neighbouring levels with increasing nonadiabatic coupling.
This formation of avoided crossings in the spectrum is a clear
signature of the classical chaos in the quantum spectrum (see
e.~g.~~\cite{H}). As in Fig.~3(b) the average Bloch variables remain
localized in a band around the $x=0$ line, this band being embedded in
a broad band of variances (Fig.~4(b)).

Although our analysis demonstrates how irregularities in the spectrum
for increasing coupling strength appear, not all regular structures
have dis\-appeared. In particular, the smooth dependence of $\Delta
E_i$ on $E_i$ associated with the bonding branch of the adiabatic
potential and observed for energies below the onset of the mixing of
the adiabatic sequences, is still visible for higher energies by a
thought continuation of this line into the mixing region.  This is
clearly observed for all the coupling cases in the figs.~1(a)-3(a) and
this feature is still present even for the strong coupling of
Fig.~4(a). There, however, the lower boundary of the spacings data
has been dissolved in contrast to the previous figures.

We note that in the high coupling case of Fig.~4 the energy interval
for which correct numerical eigenvalues are available is smaller than
in figs.~2 and 3. This is due to the stronger mixing of the basis
states used for the numerical diagonalization which makes the finite
basis size perceptible at lower energies. Beside this problem we
mention that from the point of view of a spectrum formation starting
from adiabatic eigenstates the region with relevant nonadiabatic
mixing shifts to {\em higher} energies with increasing coupling.  This
is due to the steepening of the upper branch of the adiabatic
potential (2) which increases the energy of the associated states.
Correspondingly for increasing coupling one has to pass to higher
energies in order to observe the effect of the nonadiabtic mixing on
the spectrum. Hence any finite basis used in the numerical
diagonalization restricts both the accessible energy interval and the
amount of numerical data in the mixing region for the strong coupling
case.

The analysis of the eigenstates is complemented by considering 
the degree of excitation in the vibronic subspace in form of the 
average number of vibrational quanta $n_i$  associated with each 
eigenstate $n_i  = \sum_{\nu,s}\nu |c_{\nu,s}^{(i)}|^2$. 
The corresponding data for $n_i$  and 
its variances as a function of the energy are shown in  
figs.~5(a)-(d) for the same parameters as used before. In Fig.~5(a) 
the number $n_i$  and its variance are displayed for the low coupling 
case corresponding to Fig.~1. As for the Bloch variables in 
Fig.~1(b) the resolution into two well resolved strands is 
clearly seen. The strands demonstrate the splitting of the degree 
of vibrational excitation of the eigenstates into well separated 
high and low energy parts corresponding to vibronic wave functions 
with a large and small number of nodes, respectively. The 
presence of two well resolved species of states in the two site 
excitonic~- vibronic system for certain parameter values and 
their possible experimental consequences were reported in a 
number of recent papers (see \cite{W} and references therein). We 
note that this feature is also obtained by the Bloch projection 
method, i.~e.~figs.~1(b) and 5(a) are in fact complementary. The 
figs.5(b)-(d) show the 
irregularities arising in the distribution of vibrational excitation 
between the eigenstates. Increasing the coupling the growing 
disorder in the spectrum is complemented by transitions between 
states with low and high levels of vibrational excitation
as is evident from figs.~5(b) and 5(c). In the level of 
vibrational excitation one now finds patterns showing 
growing complexity in the transitions between states of low and 
high vibrational excitation. We note that the pattern of 
Fig.~5(c) must be considered with some caution: in this case 
the variances are much broader than in the case of Fig.~5(b). 
Finally in the high coupling case presented in Fig.~5(d) the 
lower part of the $n$~-~distribution has become 
irregular, though there is clearly an upper limit in the level 
of vibrational excitation visible. As in Fig.~4(a) this 
structure can again be traced back to the bonding branch of the 
adiabatic potential, i.~e.~some regular features remain. 

Summarizing the result of the Bloch projection method we can 
show by this method that for the low coupling case some apparent 
irregularities in the spectrum can be resolved and the spectrum 
in fact be interpreted as an approximate superposition of 
adiabatic level sequences. For higher couplings this method 
shows that the Bloch variables merge into a band
indicating that a 
separation of the spectrum into adiabatic sequences is not 
possible any more. Considering the distribution of vibrational 
excitation between the eigenstates we find the same behaviour, 
i.~e.~both views are in fact complementary. The Bloch projection 
seems however simpler since it employs a projection on the low 
dimensional space of two bonding/antibonding states. The Bloch 
projection allows to follow how the genuine irregularities in the 
spectrum appear together with a destruction of the classification 
of eigenstates such as bonding and antibonding according to 
criteria other than their energy. An essential point is that for 
the coupled excitonic~- vibronic system the formation of 
irregularities in spectrum and associated eigenstates is not 
complete: some of the regular structures are conserved. This 
feature justifies the classification of this situation as 
incipience of quantum chaos.

\section{Conclusions}
\begin{enumerate}
\item The main aim of this paper was to show how irregularities in 
the spectrum and associated eigenvectors of the fully quantized 
electronic two site system coupled to a vibration appear when a 
characteristic coupling parameter is increased. The considered 
system belongs to the class of models which show regular and 
chaotic dynamics in the mixed quantum~- classical description 
when a stepwise quantization is applied treating the electronic 
subsystem in the quantum but the vibronic subsystem in the 
classical context \cite{ES}. Hence it is natural to ask for the 
"fingerprints" of the dynamical chaos in the mixed description 
after full quantization is performed. Considering the fully 
quantized system we employed the adiabatic approximation as a 
standard way to define an integrable reference system and used 
the nonadiabatic couplings as the parameter to control the degree 
of nonintegrability and resulting spectral irregularities. 
In agreement with earlier findings \cite{GH,K} for this type of 
system our results show that spectral irregularities and in 
particular avoided crossings, which are a signature of dynamical 
chaos, indeed appear, but do not develop into fluctuations of the 
kind one would expect from the standard examples of quantum chaos 
when quantizing a classically chaotic system results in spectral 
fluctuations characterized by universal statistics and random 
matrix theory. Increasing the nonadiabatic couplings we rather 
find a transition like behaviour characterized by the birth of 
spectral irregularities and growing disorder in the 
characterization of eigenstates. This growing disorder in the 
eigenstates proceeds through the destruction of regular 
patterns and shows how distinction criteria of eigenstates 
other than their energy such as provided by the Bloch projection 
or the level of vibrational excitation are lost.

\item Referring to the specific way in which the spectral structures 
associated with the Born~- Oppenheimer approximation are 
dissolved when the characteristic coupling~- the nonadiabatic 
mixing of the integrable oscillators of this approximation~- is 
increased, we find that the mixture of both residual regular 
structure and disordered parts makes the spectrum in a sense 
unique. Even for large coupling the spectrum still contains the 
regular spacing dependence rooted in the adiabatic approximation 
and irregular parts scattered around it. In our view this 
situation can be best described as incipience of quantum chaos, 
which is in line with recent findings for the same kind of system 
viewed from spin~- boson coupled systems in an another parameter 
region \cite{CCM}. Our results show the specific way in which this 
incipience of quantum chaos develops by the nonadiabatic coupling 
of the integrable reference oscillators of the adiabatic 
approximation.

\item Finally we comment on the reason why the present system is of 
the incipience type thereby pointing to other more complex 
situations when the incipience case can be expected to pass 
into the situation of fully developed spectral fluctuations after 
complete quantization. From the view of the adiabatic 
approximation, there are obviously two points to mention in order 
to obtain a more complex system: i) to increase the number of 
coupled adiabatic reference oscillators, i. e. the dimension of 
the Hilbert space of the fast subsystem and ii) to increase the 
degree of nonlinearity of the reference oscillators which are 
coupled. As viewed from these points the present system is 
obviously of the simplest form, there are two basis states of the 
fast subsystem and the degree of nonlinearity of the adiabatic 
potentials (2), (3) is mild. We conclude that the increase in the 
number of coupled adiabatic reference systems and/or the degree 
of their nonlinearity is essential to further investigate the 
connection between chaos of systems treated in the mixed quantum 
- classical description of the first step and the signatures of 
this chaos after full quantization is performed in the second 
step of the Born~- Oppenheimer procedure.
\end{enumerate}

\section*{Acknowledgement}
Financial support from the Deutsche Forschungsgemeinschaft (DFG) 
is gratefully acknowledged.

\pagebreak

\pagestyle{empty}

\begin{figure}[htb]
\vspace*{-15mm}
\psfig{figure=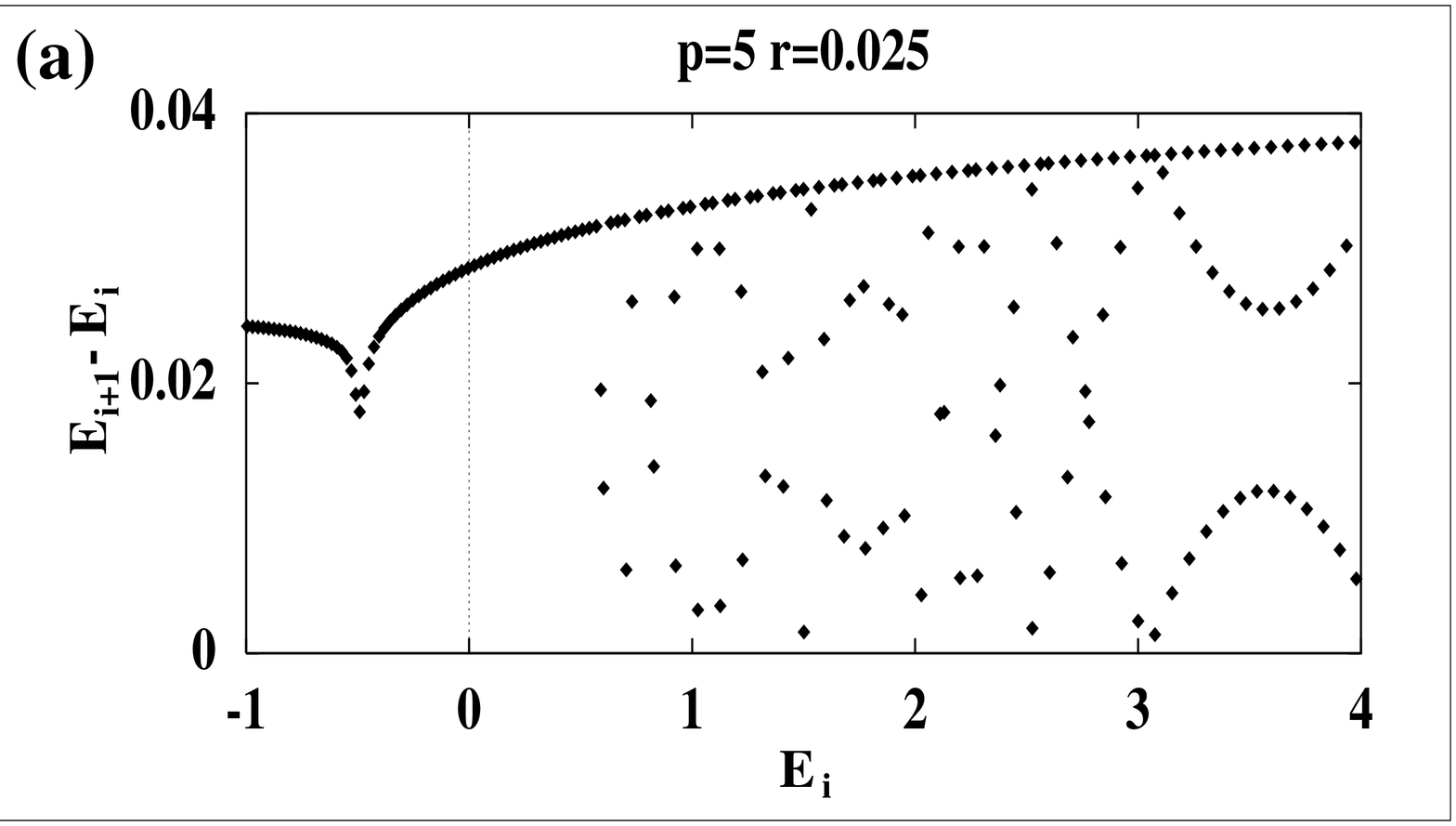,width=12cm,height=6cm,angle=-90}
\psfig{figure=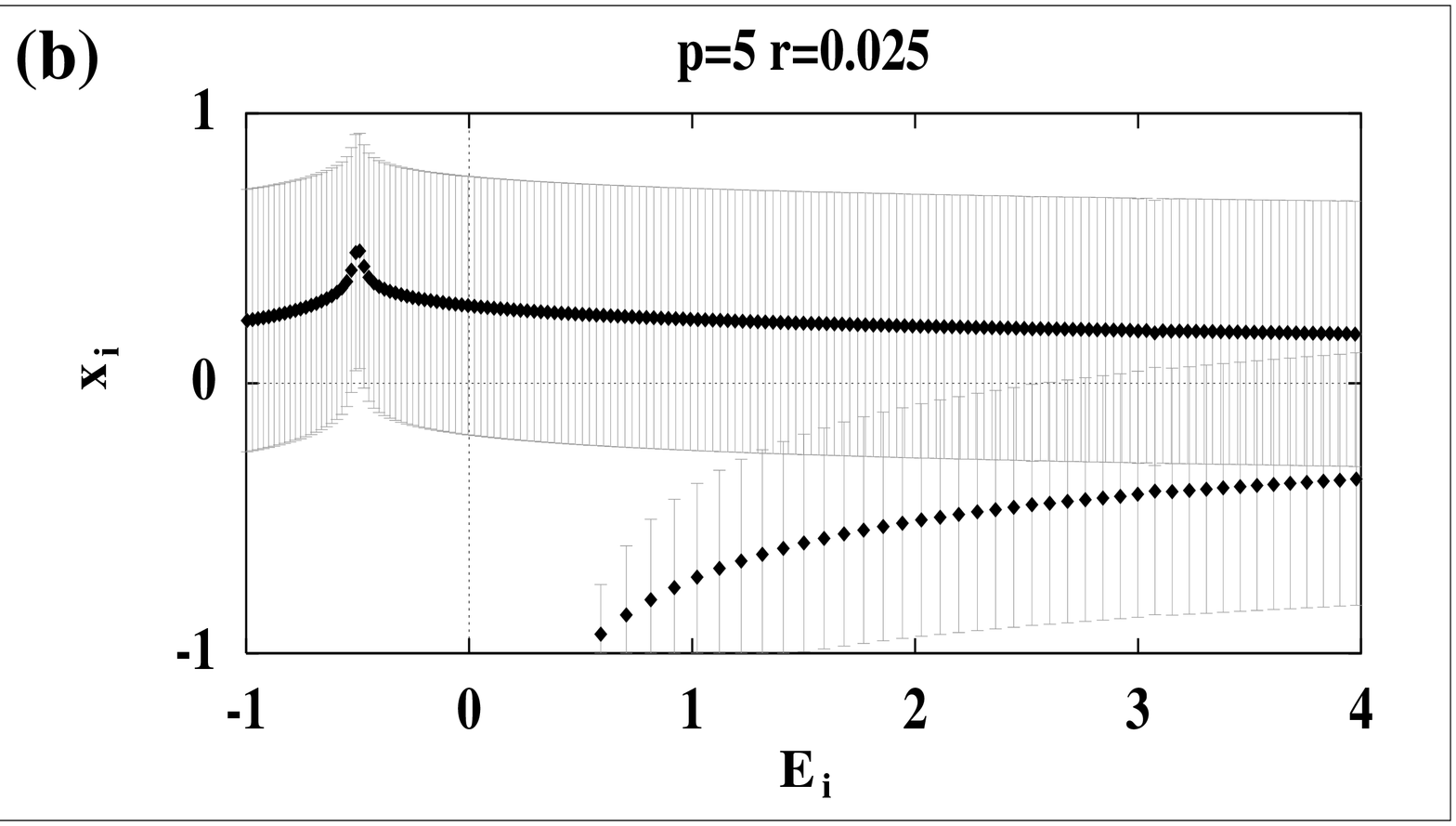,width=12cm,height=6cm,angle=-90}
\psfig{figure=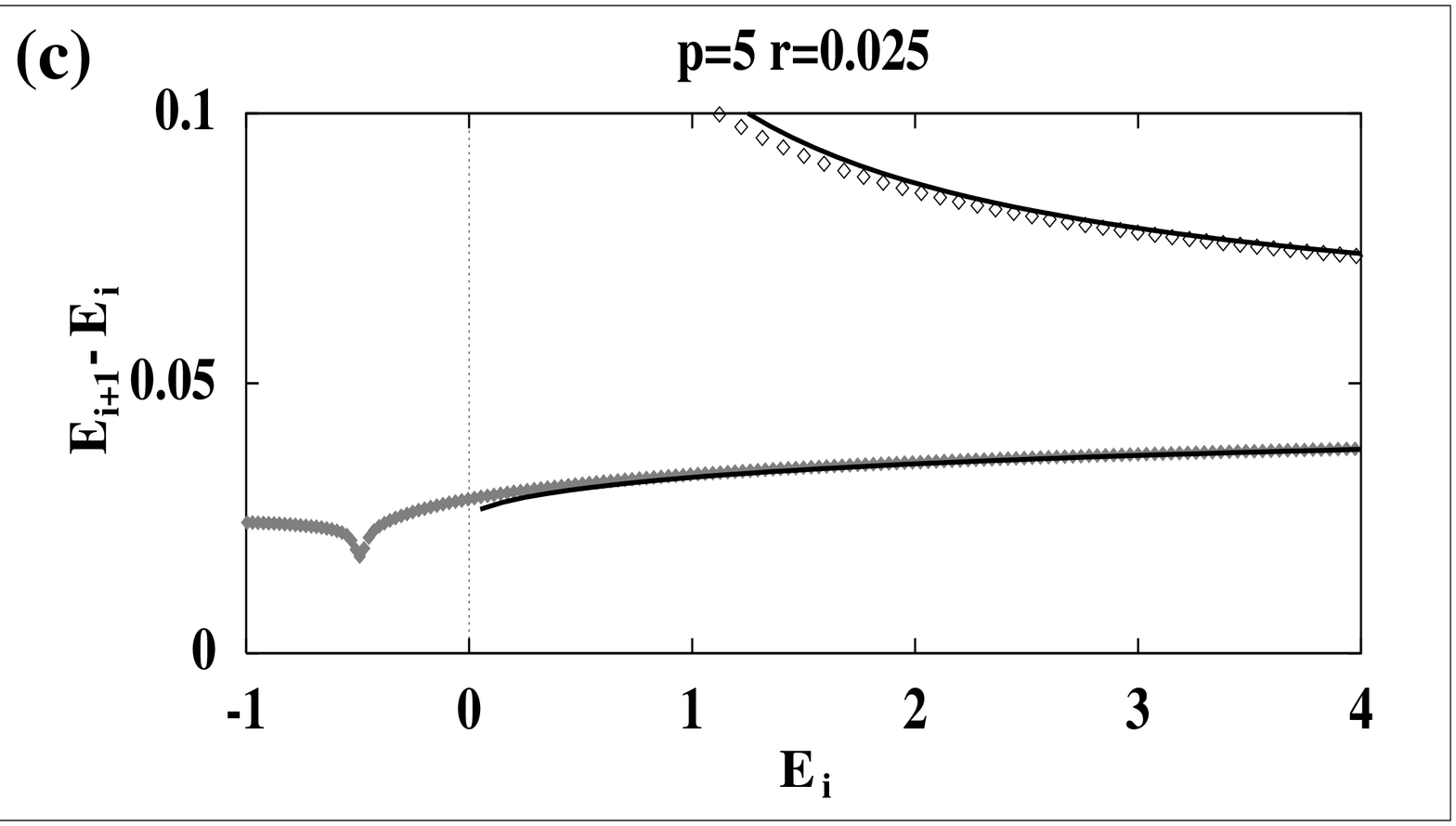,width=12cm,height=6cm,angle=-90}
\caption{Analysis of the spectrum for positive parity states
  in the low coupling case $p=5$ and $r=0.025$, all energies are in
  units of $2V$: (a) Spacings $\Delta E_i$ between adjacent levels as
  a function of $E_i$ (b) Expectation values (diamonds) and variances
  (bars) of the Bloch variable $x_i$ associated with each eigenstate.
  Two separate strands can be resolved. (c) Resolution of the spacings
  shown in (a) according to the sign of the Bloch variable $x_i$ into
  two strands. The spacings between levels with $x_i > 0$ and $x_i <
  0$ are shown with full and empty diamonds, respectively.  The
  scatter of $\Delta E_i$ values in the region of overlap of the
  adiabatic potentials is completely resolved into two monotonic
  dependences close to the level spacings of the adiabatic
  potentials (4) which are shown as full lines.
}
\end{figure}

\begin{figure}[htbp]
\thispagestyle{empty}
\psfig{figure=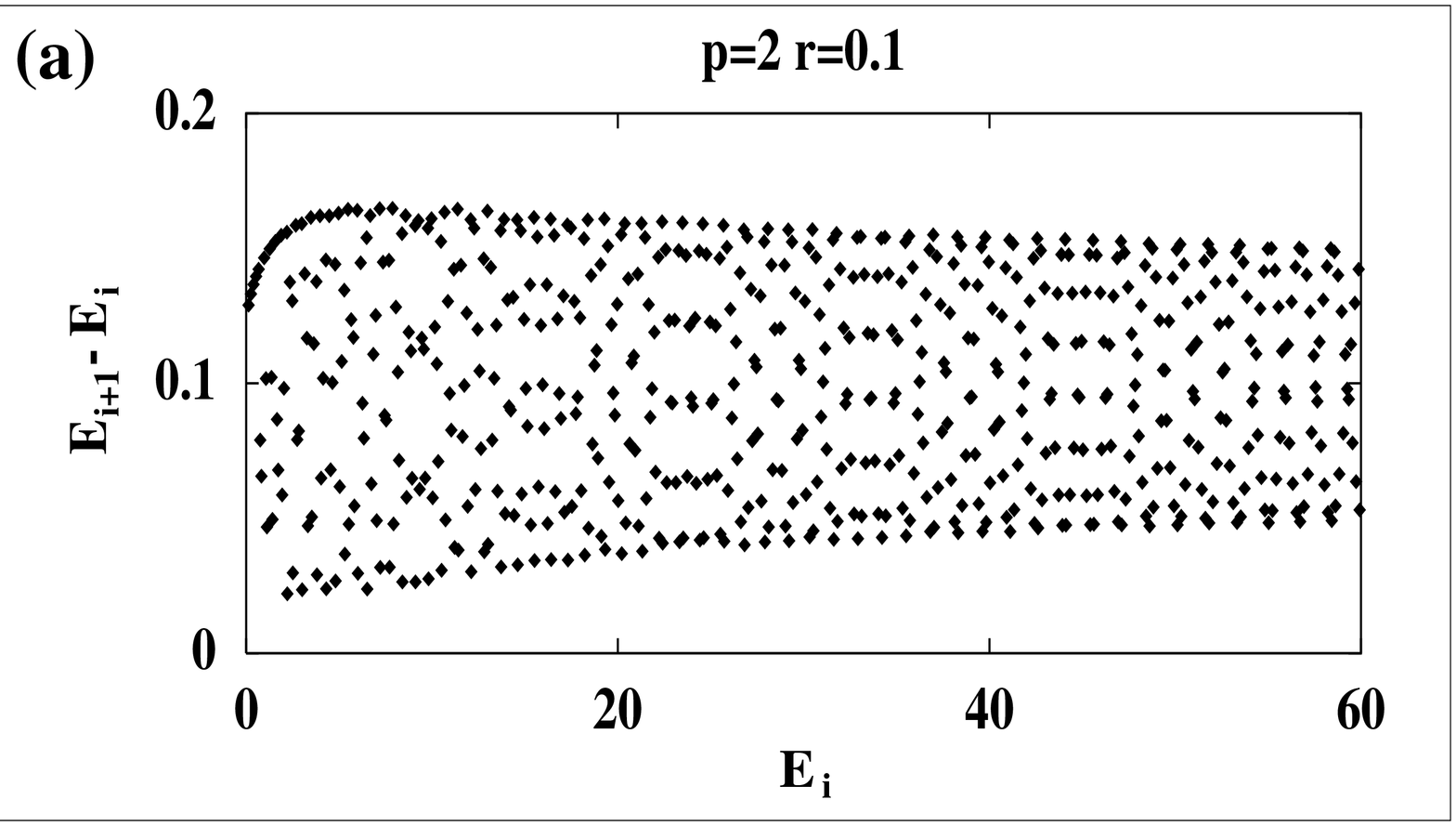,width=12cm,height=6cm,angle=-90}
\psfig{figure=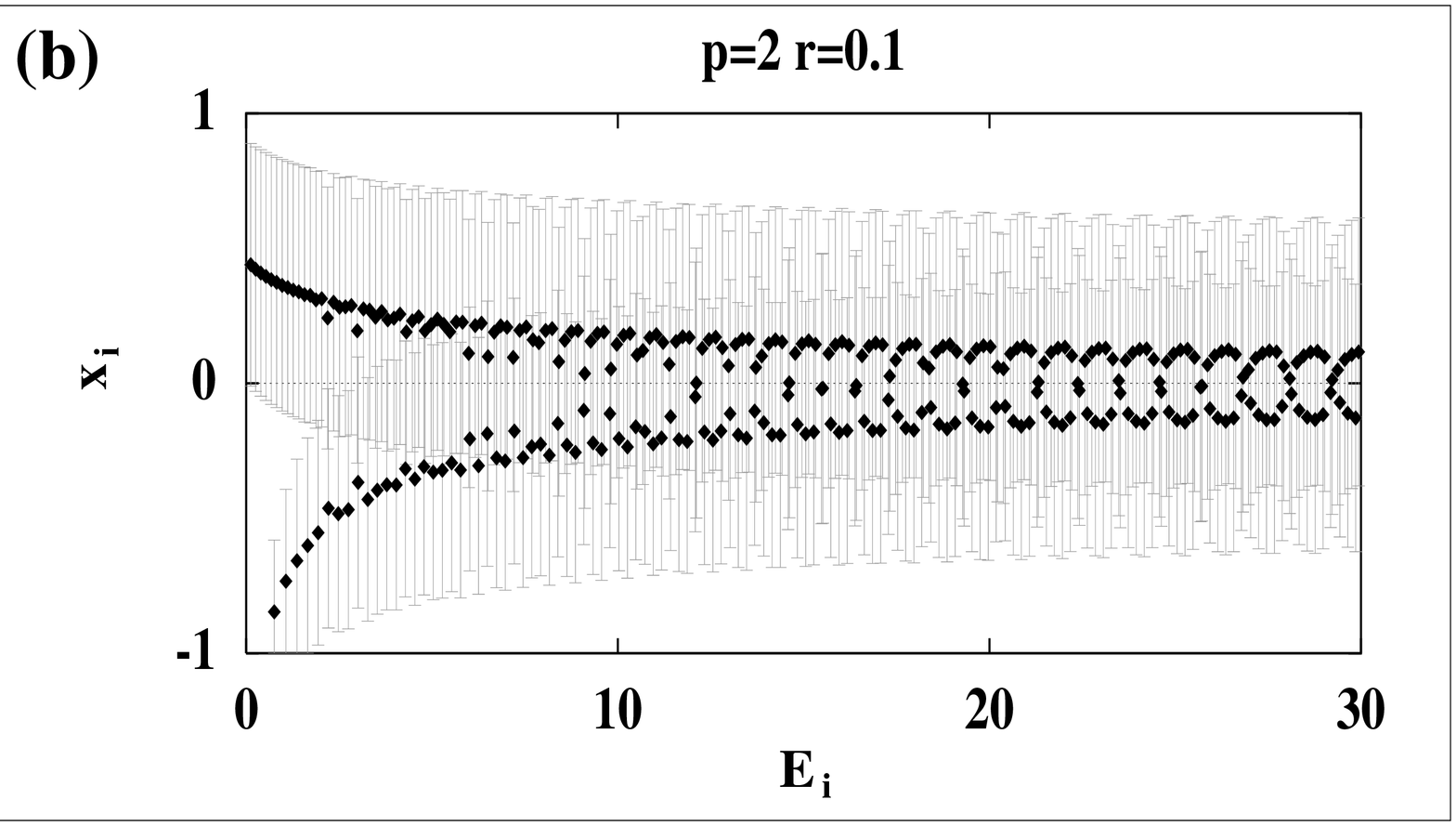,width=12cm,height=6cm,angle=-90}
\psfig{figure=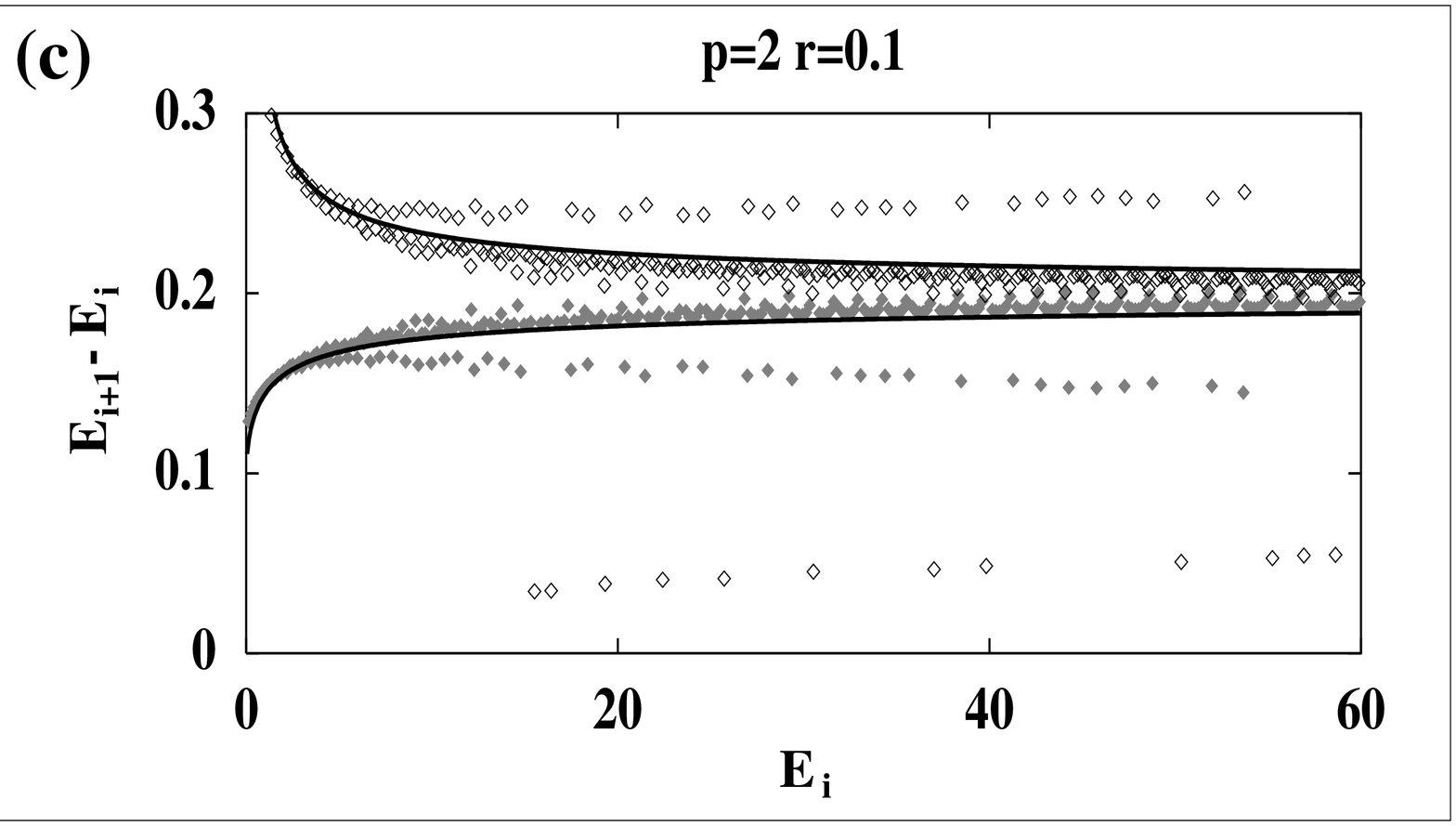,width=12cm,height=6cm,angle=-90}
\caption{
Analysis of the spectrum for $p=2$ and $r=0.1$ as in
  Fig.~1. (c) The majority of spacings for large $E$ are close to the
  full lines which mark the adiabatic level spacings (4), but others
  are far off. The parameters chosen are in the region of the gradual
  breakdown of the adiabatic approximation and mark the limit of
  applicability of the Bloch projection method for a separation of the
  spectrum into strands.}
\end{figure}

\begin{figure}[htbp]
\psfig{figure=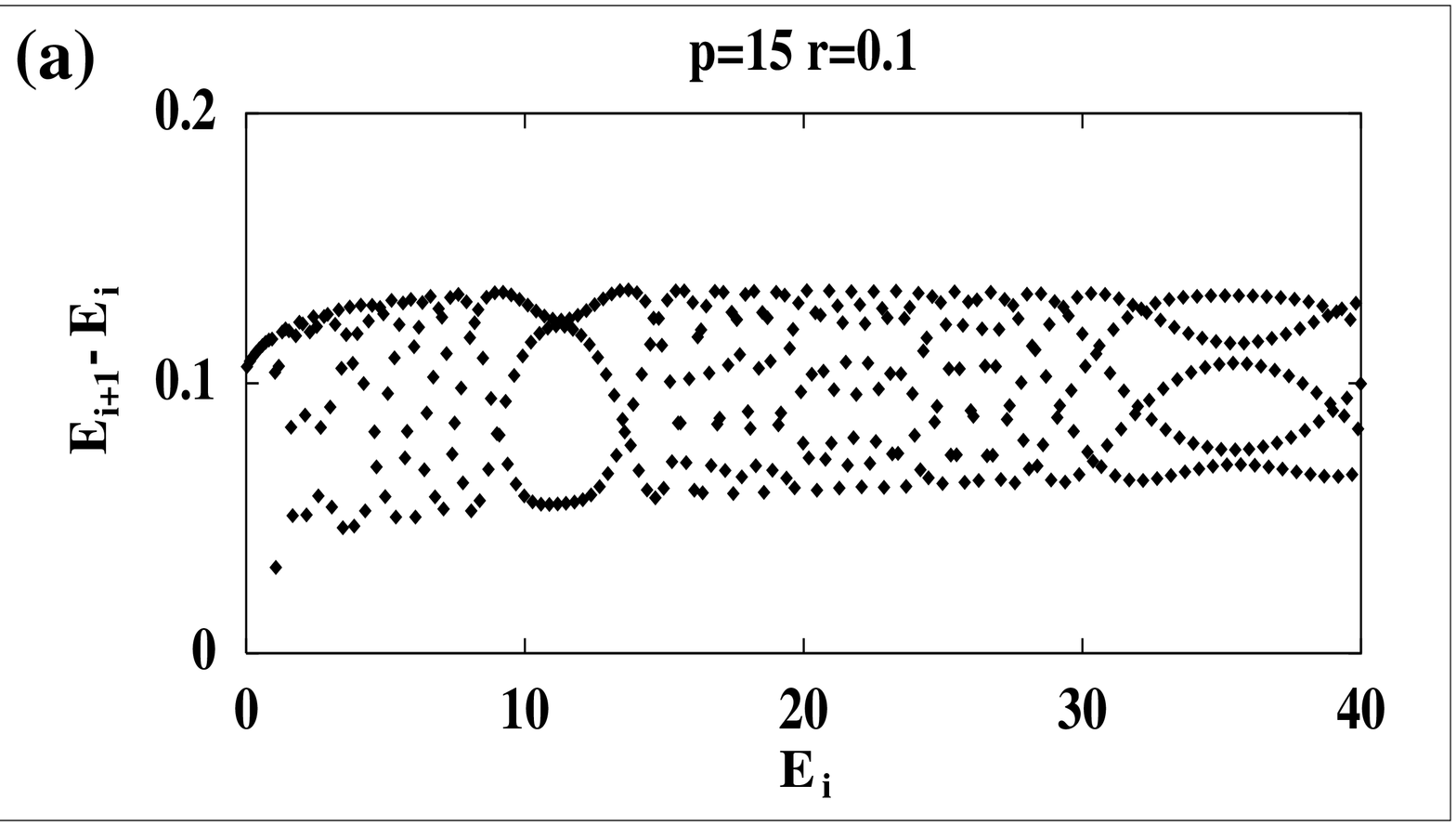,width=12cm,height=6cm,angle=-90}
\psfig{figure=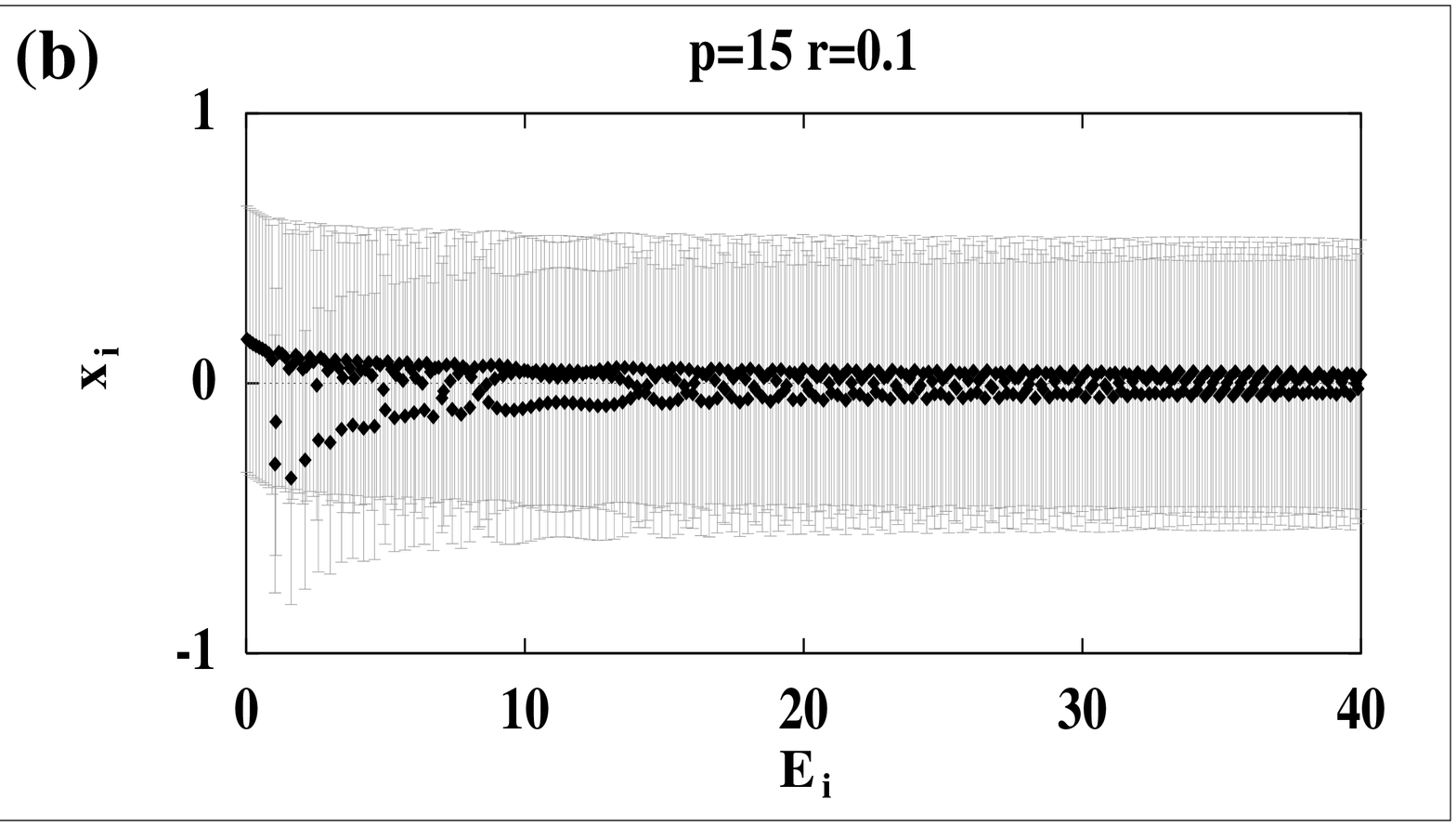,width=12cm,height=6cm,angle=-90}
\caption{
Analysis of the spectrum for $p=15$ and $r=0.1$ as in
  parts (a) and (b) of Fig.~(1).}
\end{figure}

\begin{figure}[htbp]
\psfig{figure=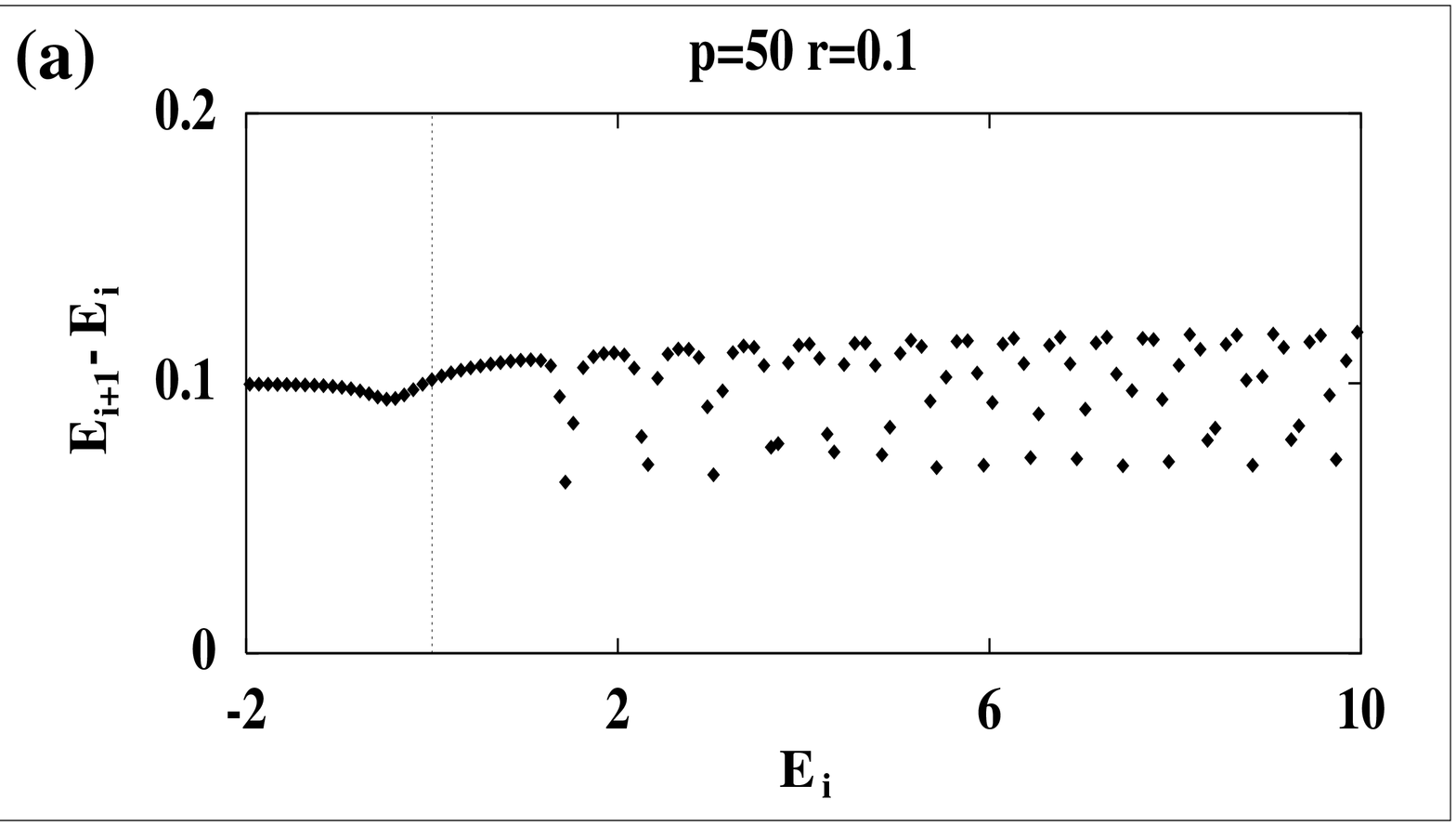,width=12cm,height=6cm,angle=-90}
\psfig{figure=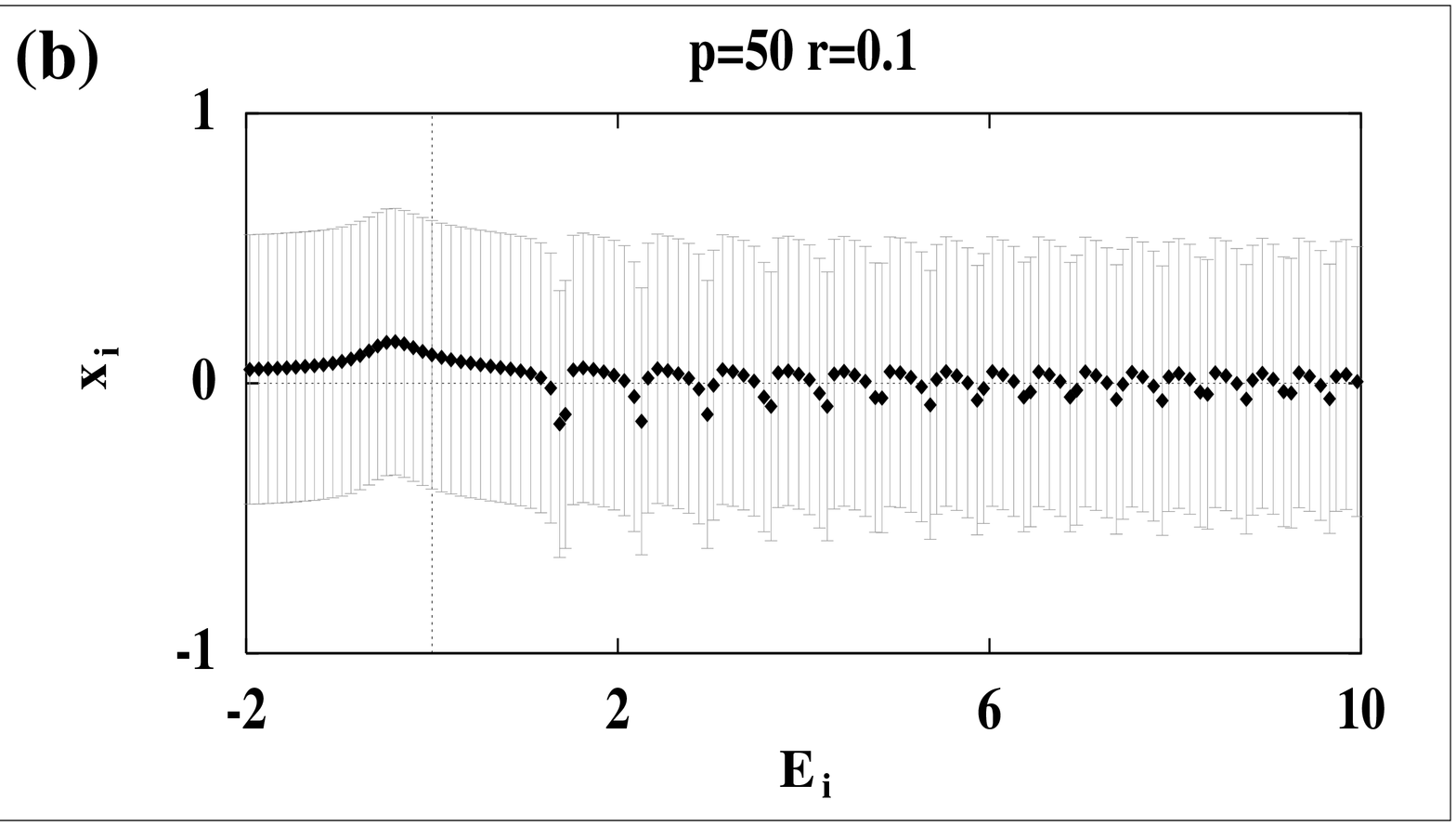,width=12cm,height=6cm,angle=-90}
\caption{
Analysis of the spectrum for $p=50$ and $r=0.1$ as in
  parts (a) and (b) of Fig.~(1).}
\end{figure}

\begin{figure}[htbp]
\begin{center}
\begin{tabular}{cc}
\hspace*{-1cm}
\psfig{figure=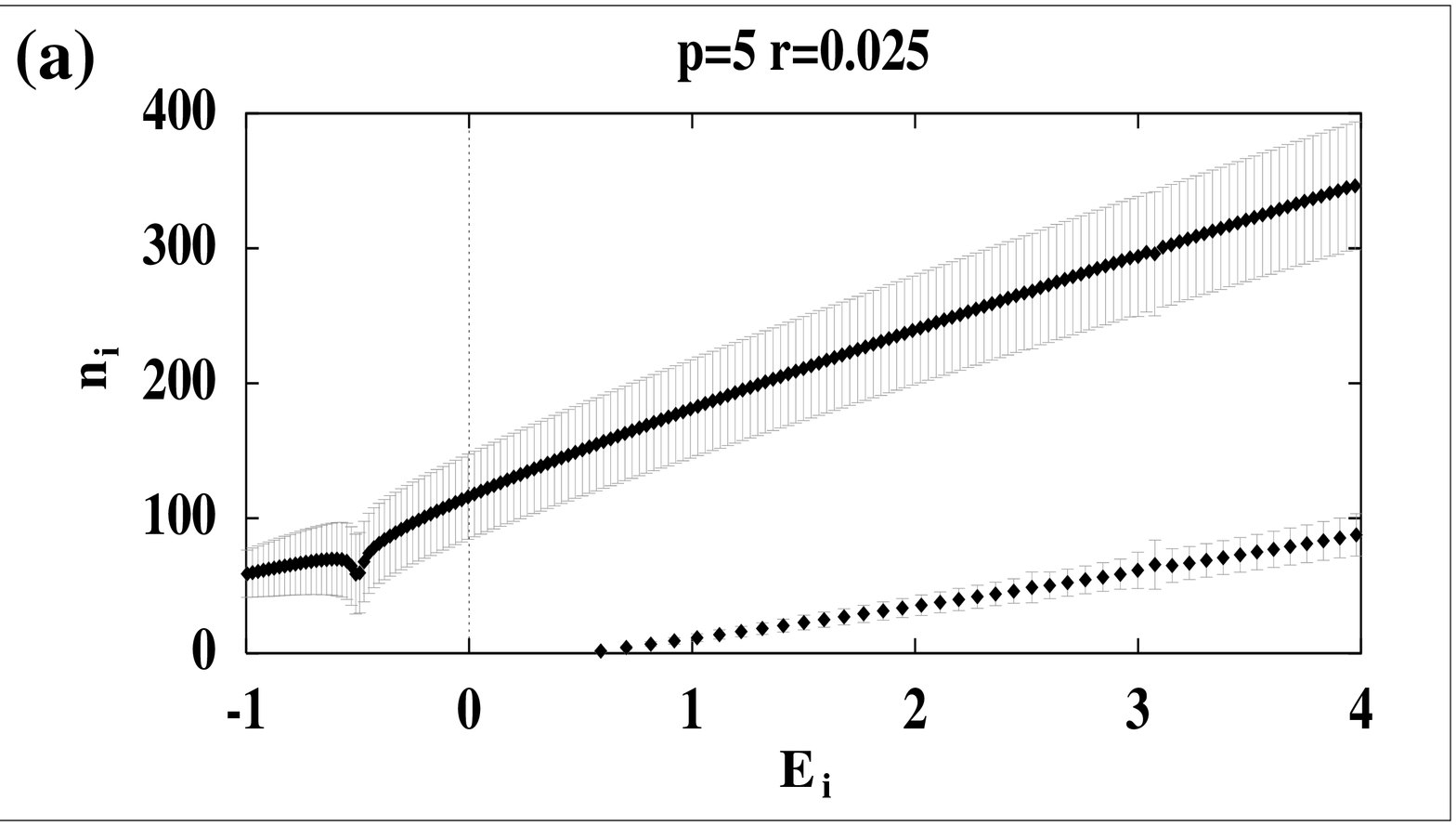,width=7cm,height=4cm,angle=-90}&
\psfig{figure=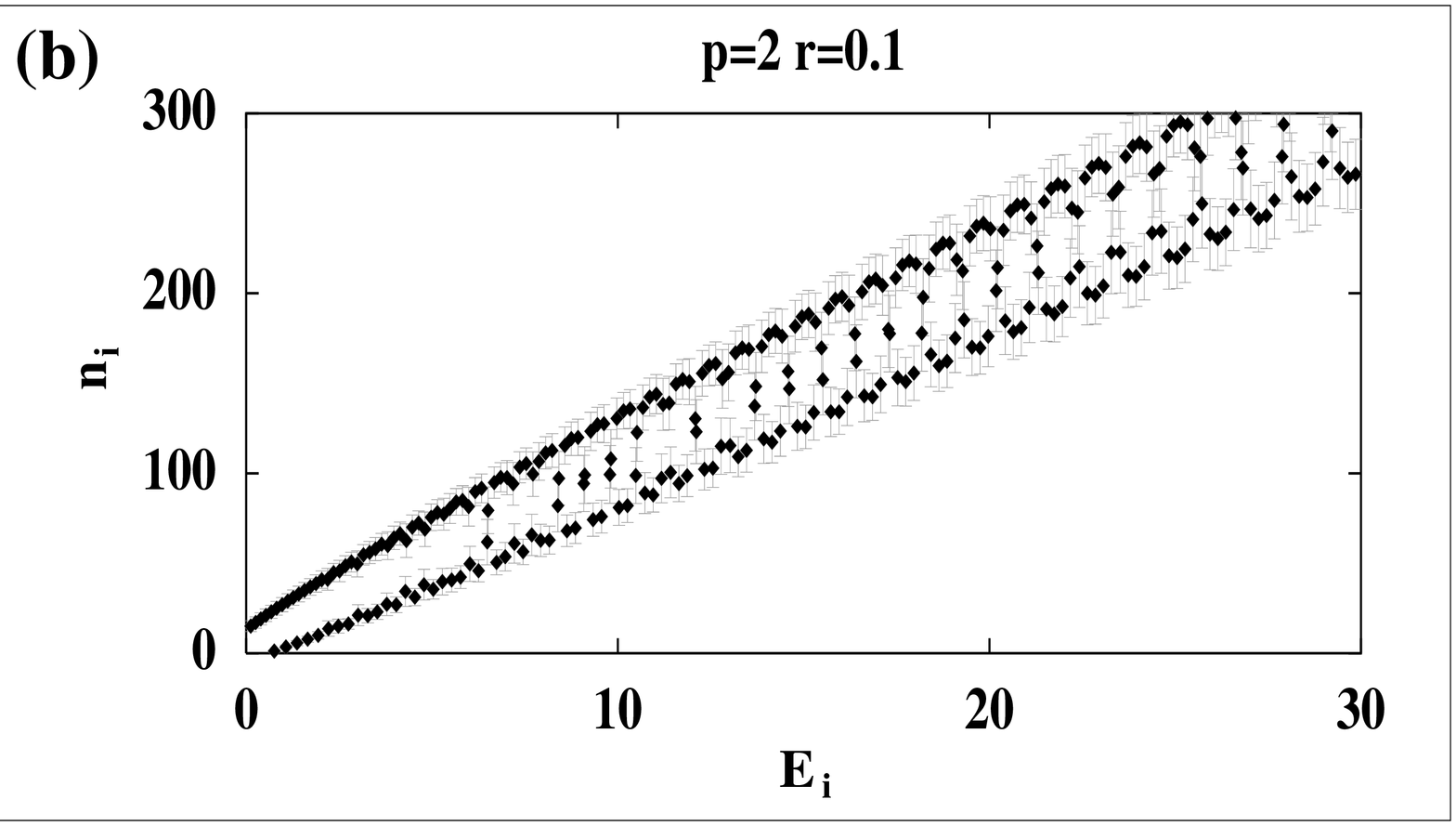,width=7cm,height=4cm,angle=-90}\\
\hspace*{-1cm}
\psfig{figure=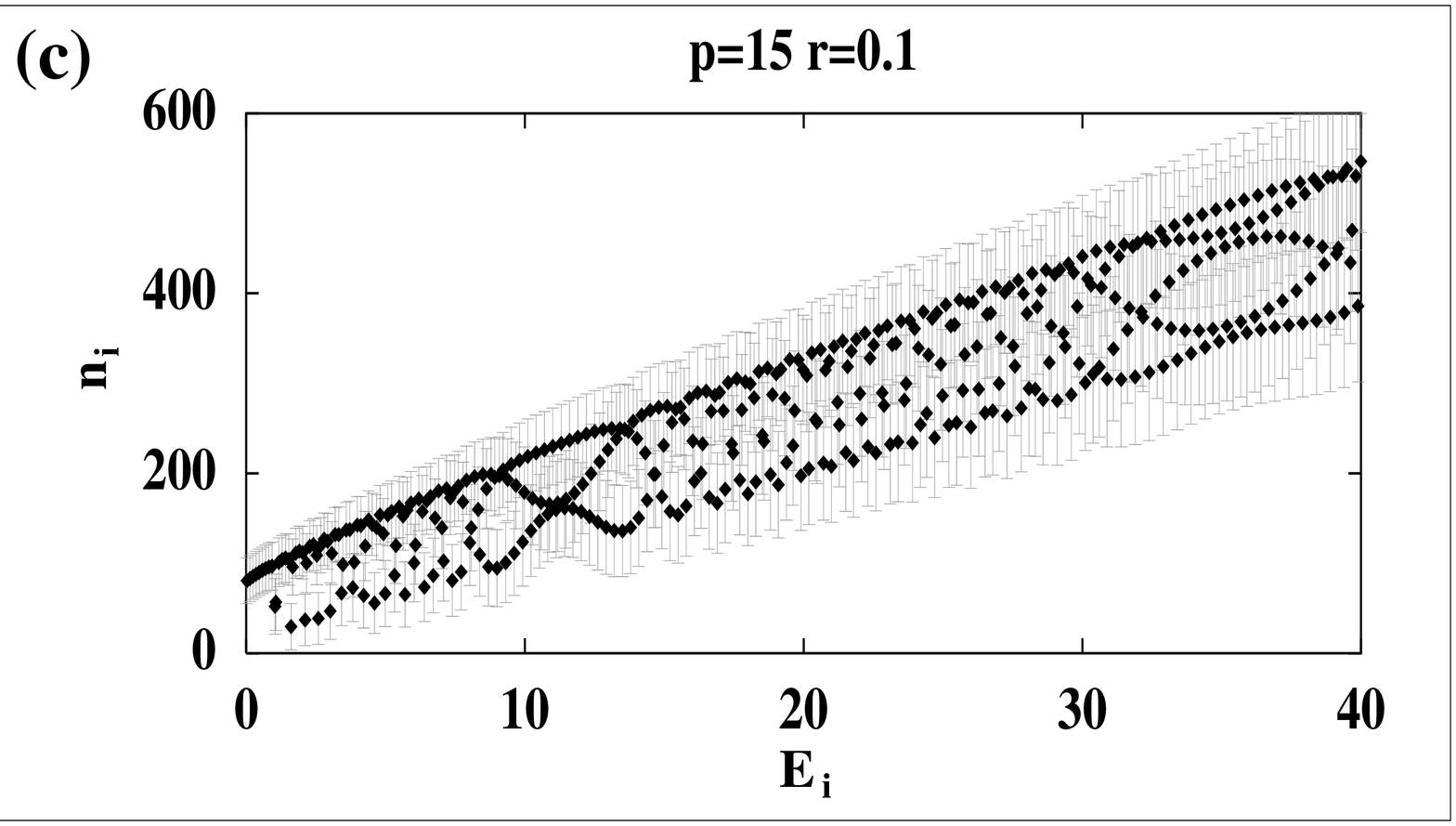,width=7cm,height=4cm,angle=-90}&
\psfig{figure=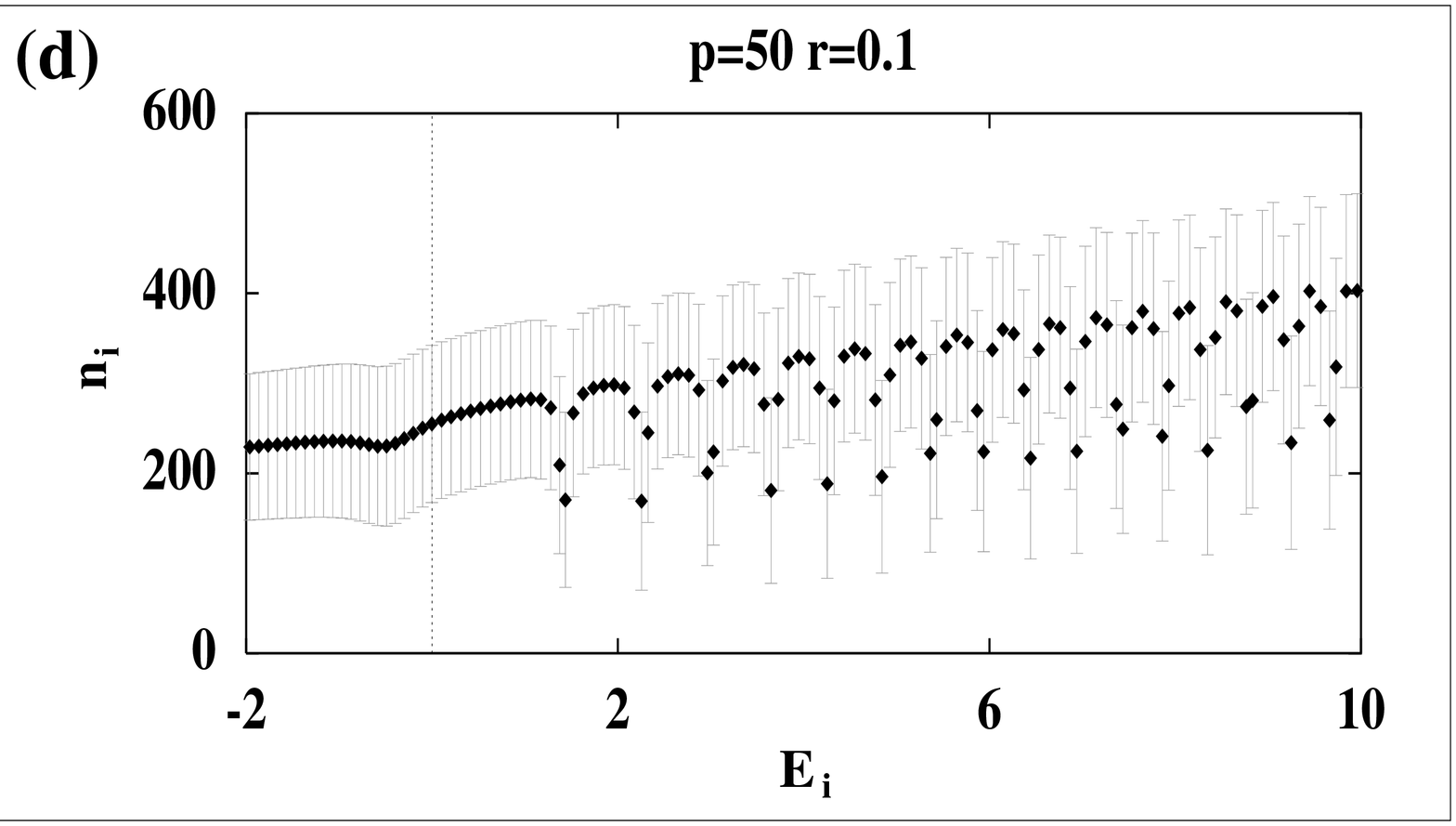,width=7cm,height=4cm,angle=-90}
\end{tabular}
\end{center}
\caption{
Expectation values (diamonds) and variances (bars) of
  the degree of vibronic excitation $n_i$ associated with each
  eigenstate. The parameters of part (a)-(d) correspond to those of
  figs.~1-4, respectively, and are given above the plots.}
\end{figure}

\end{document}